\documentclass[twocolumn]{bmcart}% uncomment this for twocolumn layout and comment line below
\usepackage{eurosym}
\usepackage{amsmath,graphicx}
\usepackage{lipsum}
\usepackage{subfigure}
\usepackage{setspace}
\usepackage{epstopdf}
\usepackage{bigints}

\usepackage{xcolor,soul}
\usepackage[normalem]{ulem}

\usepackage{array}

\usepackage{amsthm,amsmath}

\usepackage[utf8]{inputenc} %unicode support

\usepackage{booktabs}
\usepackage{bm}
\usepackage{amsfonts}
\usepackage{amssymb}
\usepackage{algorithm}
\usepackage{algpseudocode}
\usepackage{tikz}
\usetikzlibrary{arrows}
\usepackage{verbatim}
\usetikzlibrary{bayesnet}
\usepackage{color,soul}

\usepackage{graphicx}
\def\tce{\textcolor{black}}

\usepackage{pgf}
\usetikzlibrary{arrows,automata}
  
\tikzset{
    nand/.style = {-|},
    and/.style = {->}
}

\def\Piz{{\bf \Pi}_{0|0}}
\def\Pik{{\bf \Pi}_{k|k}}

\def\Pikmm{{\bf \Pi}_{k-1|k-1}}
\def\n{{\bf n}}
\def\bt{\boldsymbol{\beta}}
\def\v{{\bf{v}}}
\newcommand{\bal}{\begin{aligned}}
\newcommand{\eal}{\end{aligned}}
\newcommand{\beq}{\begin{equation}}
\newcommand{\eeq}{\end{equation}}
\def\x{{\bf x}}

\def\Xk{{\bf X}_k}
\def\Yk{{\bf Y}_k}
\def\Y{{\bf Y}}
\def\bY{\mathbb{Y}}
\def\sY{\mathcal{Y}}
\def\X{{\bf X}}

\def\nk{{\bf n}_k}
\def\f{{\bf f}}
\def\N{\mathcal{N}}
\startlocaldefs

\endlocaldefs

\usepackage{multicol}
\usepackage{lipsum}
\begin{document}

%%% Start of article front matter
\begin{frontmatter}

\begin{fmbox}
\dochead{Research}


\title{Scalable optimal Bayesian classification of single-cell trajectories under regulatory model uncertainty}

%%%%%%%%%%%%%%%%%%%%%%%%%%%%%%%%%%%%%%%%%%%%%%
%%                                          %%
%% Enter the authors here                   %%
%%                                          %%
%% Specify information, if available,       %%
%% in the form:                             %%
%%   <key>={<id1>,<id2>}                    %%
%%   <key>=                                 %%
%% Comment or delete the keys which are     %%
%% not used. Repeat \author command as much %%
%% as required.                             %%
%%                                          %%
%%%%%%%%%%%%%%%%%%%%%%%%%%%%%%%%%%%%%%%%%%%%%%

\author[
   addressref={af1}%,                   % id's of addresses, 
%   email={ehsanr@tamu.edu}   % email address
]{\inits{EH}\fnm{Ehsan} \snm{Hajiramezanali}}
\author[
   addressref={af1}
  % noteref={n1},                       
  % email={m.imani88@tamu.edu}
]{\inits{MI}\fnm{Mahdi} \snm{Imani}}
\author[
   addressref={af1}
 %  email={ulisses@tamu.edu}
]{\inits{UBN}\fnm{Ulisses} \snm{Braga-Neto}}
\author[
    corref={af1}, 
   addressref={af1},
   email={xqian@ece.tamu.edu}
]{\inits{XQ}\fnm{Xiaoning} \snm{Qian}}
\author[
   addressref={af1},
   corref={af1},
   email={edward@ece.tamu.edu}
]{\inits{ERD}\fnm{Edward R} \snm{Dougherty}}

%%%%%%%%%%%%%%%%%%%%%%%%%%%%%%%%%%%%%%%%%%%%%%
%%                                          %%
%% Enter the authors' addresses here        %%
%%                                          %%
%% Repeat \address commands as much as      %%
%% required.                                %%
%%                                          %%
%%%%%%%%%%%%%%%%%%%%%%%%%%%%%%%%%%%%%%%%%%%%%%

\address[id=af1]{%                           % unique id
  \orgname{Department of Electrical and Computer Engineering, Texas A\&M University}, % university, etc
  \street{MS3128 TAMU},                     %
  \postcode{77843}                                % post or zip code
  \city{College Station, TX},                              % city
  \cny{USA}                                    % country
}

%%%%%%%%%%%%%%%%%%%%%%%%%%%%%%%%%%%%%%%%%%%%%%
%%                                          %%
%% Enter short notes here                   %%
%%                                          %%
%% Short notes will be after addresses      %%
%% on first page.                           %%
%%                                          %%
%%%%%%%%%%%%%%%%%%%%%%%%%%%%%%%%%%%%%%%%%%%%%%

\begin{artnotes}
%\note{Sample of title note}     % note to the article
%\note[id=n1]{Equal contributor} % note, connected to author
\end{artnotes}

%\end{fmbox}% comment this for two column layout

%%%%%%%%%%%%%%%%%%%%%%%%%%%%%%%%%%%%%%%%%%%%%%
%%                                          %%
%% The Abstract begins here                 %%
%%                                          %%
%% Please refer to the Instructions for     %%
%% authors on http://www.biomedcentral.com  %%
%% and include the section headings         %%
%% accordingly for your article type.       %%
%%                                          %%
%%%%%%%%%%%%%%%%%%%%%%%%%%%%%%%%%%%%%%%%%%%%%%

\begin{abstractbox}

\begin{abstract} % abstract
\parttitle{Background} %if any
Single-cell gene expression measurements offer opportunities in deriving mechanistic understanding of complex diseases, including cancer. However, due to the complex regulatory machinery of the cell, gene regulatory network (GRN) model inference based on such data still manifests significant uncertainty. 
\parttitle{Results} %if any
The goal of this paper is to develop optimal classification of single-cell trajectories accounting for potential model uncertainty.
Partially-observed Boolean dynamical systems (POBDS) are used for modeling gene regulatory networks observed through noisy gene-expression data. 
We derive the exact optimal Bayesian classifier (OBC) for binary classification of single-cell trajectories. The application of the OBC becomes impractical for large GRNs, due to computational and memory requirements.
To address this, we introduce a particle-based single-cell classification method that is highly scalable for large GRNs with much lower complexity than the optimal solution. 
\parttitle{Conclusion}
The performance of the proposed particle-based method is demonstrated through numerical experiments using a POBDS model of the well-known T-cell large granular lymphocyte (T-LGL) leukemia network with noisy time-series gene-expression data.
\end{abstract}

%%%%%%%%%%%%%%%%%%%%%%%%%%%%%%%%%%%%%%%%%%%%%%
%%                                          %%
%% The keywords begin here                  %%
%%                                          %%
%% Put each keyword in separate \kwd{}.     %%
%%                                          %%
%%%%%%%%%%%%%%%%%%%%%%%%%%%%%%%%%%%%%%%%%%%%%%


% MSC classifications codes, if any
%\begin{keyword}[class=AMS]
%\kwd[Primary ]{}
%\kwd{}
%\kwd[; secondary ]{}
%\end{keyword}

\end{abstractbox}
\end{fmbox}% uncomment this for twcolumn layout

\end{frontmatter}

\section*{Background}

A key issue in genomic signal processing is to classify normal versus
cancerous cells, different stages of tumor development,
or different prospective drug response. 
Previous gene-expression technologies, such as microarray and RNA-Seq, typically measure the average
behavior of tens of thousands of cells \cite{hajiramezanali2018differential,hajiramezanali2018bayesian,li2017asxl1}. By contrast, the recent advances in
next-generation sequencing technologies have allowed in-depth
investigation of the transcriptome at a single-cell resolution~\cite{lake2018integrative,fiers2018mapping}. 

Gene regulatory networks (GRNs) govern the functioning of key cellular
processes, such as stress response, DNA repair, and
other mechanisms involved in complex diseases such as cancer. 
Often, the relationship among genes can be described by logical rules updated at discrete time
intervals with each gene have Boolean states: 0
(OFF) or 1 (ON)~\cite{shmulevich2014genomic}. The Partially-Observed Boolean dynamical system (POBDS) model
\cite{ImanBrag:TSP,ImanBrag:TCST,ImanBrag:LQR} is a rich framework for modeling the behavior of GRNs observed through contemporary gene-expression technologies, as it allows indirect and incomplete observation of gene states.
Several tools for the
POBDS model have been developed in recent years,
such as the optimal filter and smoother based on
the Minimum Mean Square Error (MMSE) criterion, termed as the Boolean Kalman filter
(BKF) and Boolean Kalman smoother (BKS)~\cite{ImanBrag:TSP}, respectively.

In~\cite{Alireza:TCBB1} and~\cite{Alireza:TCBB2}, the maximum-likelihood (ML) based classification of single-cell trajectories has been developed. The method uses the ML-adaptive filter proposed in~\cite{ImanBrag:TSP} for estimation of the unknown parameters, followed by the Bayes classifier tuned to the ML parameter estimates. The drawback of this method is its inability to use prior knowledge in deriving the classifier. 
In~\cite{Alireza:BMC}, the intrinsically
Bayesian robust (IBR) classifier for the trajectories is developed. This IBR classifier is optimal relative to the prior distribution of unknown parameters. 

In this paper, assuming that there are two classes, healthy ($c = 0$) and
cancerous ($c = 1$), we derive the optimal Bayesian classifier (OBC) \cite{dalton2013optimal, dalton2014intrinsically} for classification of single-cell trajectories. 
The difference between the OBC and IBR classifiers is that in the OBC the expectation of
the class-conditional densities is taken over the posterior
distribution of the unknown parameters \cite{boluki2017incorporating,boluki2017constructing}, whereas in the IBR classifier the expectation
is taken over the priors.

Despite the optimality of the developed OBC for single-cell trajectories, its exact computation for large GRNs becomes intractable, due to the large size of the matrices involved. In this paper, we develop a particle-based OBC to scale up the classification of single-cell trajectories. 
The proposed method contains a bank of Auxiliary Particle-Filter implementations of the Boolean Kalman Filter (APF-BKF) proposed in~\cite{ImanBrag:PF}, for both training and test processes. 

\tce{Our contributions are twofold: 1) Optimal Bayesian Classification: we derive the optimal Bayesian classifiers (OBC) for both single-cell gene expression trajectories and multiple-cell averaged gene expression with uncertain regulatory network prior; and 2) Scalability: the parallel particle filters together with the Monte-Carlo inference have been efficiently used to estimate the likelihood and stationary distributions, making the derived OBC scalable to larger gene regulatory networks.}

We apply the APF-BKF-based OBC to classify trajectories of the blood cancer T-cell large granular lymphocyte (T-LGL) leukemia. T-LGL leukemia is a chronic disease characterized by a clonal proliferation of cytotoxic T cells \cite{saadatpour2011dynamical}. A Boolean network model of T cell survival signaling in the context of T-LGL leukemia has been constructed by \cite{zhang2008network} through performing extensive literature search. Then the T-LGL network has been simplified by \cite{saadatpour2011dynamical}, which constructs the minimum network that preserves the attractor structure of that system. The reduced network contains $18$ genes, which has an optimal solution with a transition matrix with $2^{2\times 18}=68,719,476,736$ elements. By contrast, as we will show in our numerical experiments, the proposed APF-BKF-based method captures T-cell dynamics with only $1,000$ particles.

\section*{Methods}

\subsection*{Gene Regulatory Network Model}
Gene regulatory networks are modeled as partially-observed Boolean dynamical
systems (POBDS). The two components of the POBDS model are a
state space model that describes the evolution of the dynamics of the GRN, and an observation model for
the measurements. These two components are described
below.

\subsubsection*{GRN State Space Model}

The state process~\mbox{$\{\X_k;k =
  0,1,\ldots\}$}, where $\X_k \in \{0,1\}^n$, represents the
activation (ON)/inactivation (OFF) state for the corresponding gene across time. The state at
each discrete time is assumed to be updated through the nonlinear signal model
\beq
  \Xk \,=\, \f\left(\X_{k-1}\right)\,\oplus\, \nk \,,
\label{eq-sgnmodel}
\eeq
for $k=1,2,\ldots$, where $\f :\{0,1\}^n \rightarrow \{0,1\}^n$ is a
Boolean function called the network function, ``$\oplus$" indicates
component-wise modulo-2 addition, and $\nk \in \{0, 1\}^n$ is Boolean
transition noise at time $k$. The modulo-2 addition means that if a
bit in the noise $\n_{k}$ is $1$, the value of the corresponding gene in the
Boolean state $\X_k$ is flipped. 
In this paper, the noise process $\nk$ is
assumed to have independent components distributed as Bernoulli$(p)$,
where the parameter $p > 0$ models the noise ``intensity" --- the
closer $p$ is to $0.5$, the more chaotic the system will be, while a
value of $p$ close to zero means that the state trajectories are
nearly deterministic, being governed tightly by the network function
and perturbation process. The network
possesses a steady-state distribution $\pi^\infty$ describing its long-run behavior as:
\beq
\pi^\infty\, = \,\lim_{k\rightarrow\infty} [P(\X_k =\x^i), \dots, P(\X_k = \x^{2^n})]^T,
\eeq
where $\{\x^1,...,\x^{2^n}\}$ denotes the set of the corresponding network states in the Boolean vector representation.

\subsubsection*{Observation Model}
The data available to the experimenter is described by the 
observation process $\{\Yk; k=1,2,\ldots\}$, where
$\Yk = (\Y_k(1),\ldots,\Y_k(n))$ is a vector containing the transcript
abundance measurements at time $k$, for $k = 1,2,\ldots$. We consider a Gaussian linear model
\beq \label{eq:obs}
    \Yk \,=\, \boldsymbol{\lambda}\,+\,D\X_k + \v_k\,, k=1,2,\ldots,
\eeq
where $\v_k \sim \N(0,\sigma^2I_n)$ is an uncorrelated zero-mean Gaussian
noise vector, $\boldsymbol{\lambda}=[\lambda_1,...,\lambda_n]^T$ is a vector of baseline gene expressions corresponding to the ``zero'' state for each gene, and $D=\text{Diag}(\delta_1,...,\delta_n)$ is a diagonal matrix containing differential expression values for each gene along the diagonal (these indicate by how much the ``one'' state
of each gene is overexpressed over the ``zero'' state). 
Such a Gaussian linear model is an appropriate model for single-cell gene-expression data~\cite{wu2014quantitative,staahlberg2014workflow}.

\subsection*{Optimal Bayesian Classifier (OBC) for Single-Cell Trajectories}
Assume there are two POBDSs corresponding to the healthy
and cancerous (mutated) classes, each having $n$ genes. The difference between the healthy and mutated classes could be the over-expression or disruption of a value of single or multiple genes in the mutated case. Let $\bY^c =
\{\sY_c^{(1)},\sY_c^{(2)},...,\sY_c^{(D_c)}\}$ be the set of $D_c$ observed trajectories from class $c$, $c=0,1$. 
Let $\Theta=(\theta_1,...,\theta_M)$ be the uncertainty set of $M$ network functions containing the
unknown true network functions in (\ref{eq-sgnmodel}), indicating $M$ possible Boolean functions as $\{\f^c_{\theta_1},...,\f^c_{\theta_M}\}$ considering the regulatory model uncertainty for the class $c$. The prior probability of the model $\theta$ for the class $c$ is represented by $\pi(\theta\mid c)$, where $\sum_{i=1}^M \pi(\theta_i\mid c)=1$, for $c=0,1$. 
This uncertainty could arise due to some unknown regulations (i.e. interactions) between some genes in the pathway diagram (more information in Section~\ref{sec:NE}). We wish to derive the optimal Bayesian classifier (OBC) under uncertainty using all available data and prior knowledge.

If the feature-label distribution is unknown but belongs to an uncertainty class $\Theta$ of feature-label distributions, then we desire a classifier to minimize the expected error over the uncertainty class. 
This expected error is equivalent to the Bayesian minimum mean-square-error estimate \cite{dalton2011bayesian} given by
$\hat{\epsilon}(\psi) = E_{\theta|\bY^c,c} [\epsilon(\psi,\theta)]$, where $\epsilon(\psi, \theta)$ is the error of $\psi$ on the feature-label distribution parameterized by $\theta$ and
the expectation is taken over the posterior distribution of parameters
$\pi(\theta\mid\bY^c, c)$, $c=0, 1$.

For a given test trajectory $\sY$, the OBC, minimizing the Bayesian minimum mean-square error estimate, can be obtained as
\beq\label{eq:OBC}
\psi_{\rm OBC}(\sY)\,=\,\begin{cases}
0 & \text{if } p^0 \,E_{\theta\mid\bY^0,c=0}[p_\theta(\sY\mid c=0)]\geq \\
&\,\quad(1-p^0) \,E_{\theta\mid\bY^1,c=1}[p_\theta(\sY\mid c=1)].\\
1 & \text{otherwise}
\end{cases}
\eeq
where $p_\theta(.)$ denotes a probability density
function corresponding to parameter $\theta$,
$p^0$ is the prior probability of class $0$ and
\beq\label{eq:fT}
E_{\theta\mid\bY^c,c}[p_\theta(\sY\mid c)]\,=\,\sum_{\theta\in\Theta} p_\theta(\sY\mid c)\, \pi(\theta\mid\bY^c,c)\,,
 \eeq
for $c=0,1$. 

Derivation of (\ref{eq:fT}) requires computing the posterior distribution %$\pi(\theta\mid\bY^c,c)$, which can be written as:
\beq\label{eq:h1}
\bal
\pi(\theta\mid\bY^c,c)\,
&=\,\frac{p_\theta(\bY^c\mid c)\,\pi(\theta\mid c)}{\sum_{\theta'\in\Theta}p_{\theta'}(\bY^c\mid c)\,\pi(\theta'\mid c)}\,,
\eal
\eeq
where $\pi(\theta\mid c)$ denotes the prior probability of the corresponding network model $\theta$ for class $c$.
For an arbitrary trajectory $\tilde{\sY}$, we define the log-likelihood function associated to model $\theta$ and class $c$ by
\beq\label{eq:defLog}
L_c^\theta(\tilde{\sY})\,:=\,\log p_\theta(\tilde{\sY}\mid c)\,.
\eeq
Now, using the above definition in (\ref{eq:OBC})-(\ref{eq:h1}) leads to the following exact OBC solution:
\beq\label{eq:OBC2}
\psi_{\rm OBC}(\sY)\,=\,\begin{cases}
0 & \text{if } p^0 \,\tau^0_\Theta(\sY) \geq (1-p^0) \,\tau^1_\Theta(\sY) \\
1 & \text{otherwise}
\end{cases},
\eeq
where
\beq\label{eq:post}
\bal
\tau_\Theta^c(\sY)&=E_{\theta\mid\bY^c,c}[p_\theta(\sY\mid c)]\\
&=\sum_{\theta\in\Theta}\pi(\theta\mid\bY^c,c)\,p_\theta(\sY\mid c)\\
&=\sum_{\theta\in\Theta}\frac{p_\theta(\bY^c\mid c)\,\pi(\theta\mid c)}{\sum_{\theta'\in\Theta}p_{\theta'}(\bY^c\mid c)\,\pi(\theta'\mid c)}\,p_\theta(\sY\mid c)\\
&=\sum_{\theta\in\Theta}\frac{\exp(\sum_{d=1}^{D_c}L_c^\theta(\sY^{(d)}_c))\,\pi(\theta\mid c)}{\sum_{\theta'\in\Theta}\exp(\sum_{d=1}^{D_c}L_c^{\theta'}(\sY^{(d)}_c))\,\pi(\theta'\mid c)}\\
 &\qquad\qquad\qquad\qquad\qquad\qquad\times\exp(L_c^\theta(\sY)).
\eal
\eeq
The expectation in (\ref{eq:post}) is taken with respect to the posterior distribution of $\theta$, i.e. $\pi(\theta|\bY^c,c)$, as opposed to IBR classifier \cite{Alireza:BMC} that considers the prior distribution $\pi(\theta|c)$. Furthermore, 
$\log p_{\theta}(\bY^c\mid c)=\sum_{d=1}^{D_c} \log p(\sY_c^{(d)}\mid\theta, c)$ is used in (\ref{eq:post}) due to the independency of the training trajectories. 

As shown in equation~(\ref{eq:OBC2}), the OBC requires computing the log-likelihood functions of all training trajectories and test trajectory for both classes and $\theta\in\Theta$. Let $\{\x^1,...,\x^{2^n}\}$ denote the corresponding network states in the Boolean vector representation and $\Y_{1:T}=(\Y_1,...,\Y_T)$ be a single trajectory of length $T$. The log-likelihood function can be computed as
\beq\label{eq:loglik}
\bal
L_c^\theta(\Y_{1:T})&=\log p_\theta(\Y_{1:T}\mid c)\,\\
&=\log p_\theta(\Y_{1}\mid c)+\sum_{k=2}^T \log p_\theta(\Y_{k}\mid\Y_{1:k-1}, c)\,,
\eal
\eeq
where based on the POBDS model, 
\beq\label{eq:int1}
\bal
&p_\theta(\Y_k\mid\Y_{k-1},c)\\
&=\sum_{i=1}^{2^n} p_\theta(\Y_k\mid\X_{k}=\x^i,c)\,P_\theta(\X_k=\x^i\mid\Y_{1:k-1},c)\,\\
&=\sum_{i=1}^{2^n} p_\theta(\Y_k\mid\X_{k}=\x^i,c)\,\\
&\qquad\qquad\qquad\quad\times\sum_{j=1}^{2^n}  P_\theta(\X_k=\x^i\mid\X_{k-1}=\x^j,c)\\
&\qquad\qquad\qquad\quad\times P_\theta(\X_{k-1}=\x^j\mid\Y_{1:k-1},c).
\eal
\eeq

Define the conditional probability of any network state at time step $k$ for the model $\theta$ and class $c$ as 
\beq\label{eq:pi}
\bal
  &\Pik^{\theta,c} \,=\,\\
  &\left[ P_\theta\left(\X_k = \x^1 \mid  \Y_{1:k},c\right), \dots, P_\theta\left(\X_k = \x^{2^n} \mid  \Y_{1:k},c\right)\right]^T.
\eal
\eeq
Using (\ref{eq:int1}) and (\ref{eq:pi}), the log-likelihood function in (\ref{eq:loglik})
can be written in a compact form as~\cite{ImanBrag:ACC2017}
\beq\label{eq:loglik2}
\bal
L_c^\theta(\Y_{1:T})\,&=\, \sum_{k=1}^T\log ||T_k^{\theta,c}(\Y_k)\, M^{\theta,c}_k\,\Pikmm^{\theta,c}||_1\,,%\\
%&=\,\log p_\theta(\Y_{1}\mid c)+\sum_{k=2}^T \log p_\theta(\Y_{k}\mid\Y_{1:k-1}, c)\,,
\eal
\eeq
where $||.||_1$ denotes $L_1$ norm, indicating the summation in \eqref{eq:loglik}. $M^{\theta,c}_k$ is the transition matrix of the Markov state process corresponding to model $\theta\in\Theta$, with entries given by: 
\beq
\bal
    (M^{\theta,c}_k)_{ij} \,&=\, P_\theta(\X_k = \x^i \mid \X_{k-1} = \x^j,c) \\
    &=\, p^{||\f_\theta^c(\x^{j})\,\oplus\,\x^i||_1}\, (1-p)^{n-||\f_\theta^c(\x^{j})\,\oplus\,\x^i||_1},\\
    &\qquad\qquad\qquad\qquad\qquad\,\,\, i,j = 1,\ldots,2^n,
\eal
\eeq
with $p$ denoting the Bernoulli noise parameter. $T^{\theta,c}_k(\Y_k)$ is a diagonal matrix, called the update matrix, with the $i$th diagonal element given by % w matrix 
\beq
\bal
    &(T^{\theta,c}_k(\Y_k))_{ii} \,=\, p_\theta(\Y_k \mid \X_{k} = \x^i,c) \\
    &=\,\left(\frac{1}{\sqrt{2\pi\sigma^2}}\right)^n \exp\left(-\sum_{j=1}^n\frac{\left(\Y_k(j)-\lambda_j-\delta_j \x^i(j)\right)^2}{2\sigma^2}\right)
    ,
\eal
\eeq
for $i= 1,\ldots,2^n$, where $\delta_j$ and $\lambda_j$ are gene-expression parameters associated to the $j$th gene in class $c$ as defined in (\ref{eq:obs}). Notice that the initial distribution $\Piz^{\theta,c}=\pi^\infty_{\theta,c}$ is the steady-state distribution associated to model $\theta$ and the class $c$ defined as
\beq
\pi^\infty_{\theta,c}\,=\, \lim_{k\rightarrow\infty} [P_\theta(\X_{k} = \x^1 | c), \dots, P_\theta(\X_{k} = \x^{2^n} | c)]^T. % \:\: k \rightarrow \infty.
\eeq
This vector can be either computed exactly as introduced in \cite{Alireza:TCBB1} or approximated by creating multiple Monte-Carlo trajectories with relatively long horizons.

The posterior distribution can also be recursively computed according to the transition and update matrices as described in~\cite{ImanBrag:TSP}:
\begin{equation}
\label{eq:TMP}
\Pik^{\theta,c}\,=\frac{T_k^{\theta,c}(\Y_k)\,M_k^{\theta,c}\,\Pikmm^{\theta,c}}{||T_k^{\theta,c}(\Y_k)\,M_k^{\theta,c}\,\Pikmm^{\theta,c}||_1}\,, \:\:k=1,2,\ldots.
\end{equation}
The complexity of computing the log-likelihood function for a single trajectory of length $T$ is of order $O(2^{2n}\times T)$ due to the transition matrix involved in its computation. 
The whole process of the proposed OBC is presented in Algorithm~\ref{alg:OBC}.

\subsection*{Scalable Classification of Single-Cell Trajectories}
In the previous section, the exact solution for the optimal Bayesian classifier is introduced. 
However, for large systems with a large number of state variables, the exact computation of Algorithm~\ref{alg:OBC} becomes impractical. This is due to the large transition matrix with $2^{2n}$ elements required to compute the log-likelihoods, leading to exponential computational and memory complexity. Thus, the key here is to scale up the OBC for single-cell trajectories by reducing both computational and memory complexity when computing (\ref{eq:loglik}).

We adopt the Sequential Monte-Carlo (SMC) techniques~\cite{kantas2015particle,hajiramezanali2012maneuvering,ImaniAAAI19,hajiramezanali2012man,hajiramezanali2013stochastic,fouladi2013denoising} for estimating nonlinear state-space models, such as our POBDS here. These techniques approximate the target distribution using sample points (``particles") drawn from a proposal distribution, taking advantage of the fact that sampling from the proposal distribution is easier than from the target. This helps alleviate the high computation of the exact filter by using a finite set of Monte-Carlo samples.
In this paper, we use the {Auxiliary Particle Filter implementation of the Boolean Kalman Filter} (APF-BKF) proposed in~\cite{ImanBrag:PF} to deal with large GRNs. 
 
\begin{algorithm}[H]
\caption{Optimal Bayesian Classification of Single-Cell Trajectories}
\label{alg:OBC}
\begin{algorithmic}[1]
\Statex \underline{\textit{Training Process}}\vspace{1ex}
\State Compute the steady-state distribution
$\Piz^{\theta,c}=\pi_\theta^\infty$, $\theta\in\Theta$, $c=0,1$~\cite{Alireza:TCBB1}. \vspace{1ex} 
\For {$d\in\{1,...,D_c\}$}\vspace{.5ex}
\For {$c\in\{0,1\}$}\vspace{.5ex}
\For {$\theta\in\Theta$}\vspace{1ex}
\State ${L}_c^\theta(\sY^{(d)}_c)\leftarrow\textbf{Log-Lik}~(\theta,c,\sY^{(d)}_c,\pi_\theta^\infty)$\vspace{1ex}
\EndFor\vspace{.5ex}
\EndFor\vspace{.5ex}
\EndFor\vspace{1ex}
\Statex \underline{\textit{Test Process}}\vspace{1ex}
\For {$c\in\{0,1\}$}\vspace{.5ex}
\For {$\theta\in\Theta$}\vspace{1ex}
\State ${L}_c^\theta(\sY)\leftarrow\textbf{Log-Lik}~(\theta,c,\sY,\pi_\theta^\infty)$\vspace{1ex}
\EndFor\vspace{1ex}
\State $\tau_\Theta^c(\sY)=\sum_{\theta\in\Theta}\frac{\exp(\sum_{d=1}^{D_c}{L}_c^\theta(\sY^{(d)}_c))\,\pi(\theta\mid c)\exp({L}_c^\theta(\sY))}{\sum_{\theta'\in\Theta}\exp(\sum_{d=1}^{D_c}{L}_c^{\theta'}(\sY^{(d)}_c))\,\pi(\theta'\mid c)}$\vspace{1ex}
\EndFor\vspace{.5ex}
\State $\psi_{\rm OBC}(\sY)\,=\,\begin{cases}
0 & \text{if } p^0 \,\tau^0_\Theta(\sY) \geq (1-p^0) \,\tau^1_\Theta(\sY) \\
1 & \text{otherwise}
\end{cases}$. \vspace{1ex}
\end{algorithmic}
\begin{algorithmic}[1]
\hrule\vspace{1ex}
\Statex \hspace{-3ex}{\bf Log-Lik}~($\theta,c,\Y_{1:T},\Piz$)\vspace{1ex}
\State ${L}_{c}^\theta=0$.\vspace{1ex}
\For{$k =1,2,\ldots,$} \vspace{1ex}
\State $\bt_k^{\theta,c}\!=\!T_k^{\theta,c}(\Y_k)\, M_k^{\theta,c}\,\Pikmm^{\theta,c}.$\vspace{1ex}
\State $\Pik=\bt_k^{\theta,c}/||\bt_k^{\theta,c}||_1.$\vspace{1ex}
\State ${L}_c^\theta={L}_c^\theta+\log\! ||\bt_k^{\theta,c}||_1.$\vspace{1ex}
\EndFor\vspace{.8ex}
\Statex \hspace{-3ex} {\bf Return} (${L}_c^\theta$)
\end{algorithmic}
\end{algorithm}
 
Let $\Y_{1:T}$ be a given trajectory and the goal is to approximate the likelihood function in equation~(\ref{eq:loglik}) (i.e., $L_c^\theta(\Y_{1:T})$ for class $c\in\{0,1\}$ and $\theta\in\Theta$). Let there be $N$ total particles $\{\x_{k-1,i}\}_{i=1}^N$, with their associated  weights $\{w_{k-1,i}\}_{i=1}^N$. The particle filter allows us to produce an approximation for the elements in $\Pik^{\theta,c}$ of \eqref{eq:pi}, simply
by using the discrete support of the particles
\beq
\label{eq:PFSIR}
\bal
P_\theta(&\Xk \!\mid\! \Y_{1:k},c)\, \\
& \propto\, p_\theta(\Yk \!\mid\! \Xk,c) \sum_{i=1}^N P_\theta(\Xk\!\mid\!
\x_{k-1,i},c)\,w_{k-1,i}.
\eal
\eeq
Usually only a few particles have significant weights after a few iterations of the algorithm and most particles have negligible weights.
APF-BKF is a look-ahead method that predicts the location
of particles with a high probability at time $k$ based on the observations at time step $k-1$, with
the purpose of making the subsequent resampling step
more efficient. Without the look-ahead, the basic
algorithm blindly propagates all particles, even those in
low probability regimes.

The APF-BKF algorithm defines:
\beq
\bal
P_\theta(&\Xk,\zeta_k \!\mid\! \Y_{1:k},c)\,\\
&\propto\,
p_\theta(\Yk \!\mid\! \Xk,c)\,P_\theta(\Xk\!\mid\!
\x_{k-1,\zeta_k},c)\,w_{k-1,\zeta_k}\,, \:\: %\zeta_k = 1,\ldots,N,
%&\propto\,p_\theta(\Yk \!\mid\! \Xk,c)\,P_\theta(\Xk\!\mid\!
%\tilde{\x}_{k-1,\zeta_k})\,w_{k-1,\zeta_k}\,,
\label{eq:apfd}
\eal
\eeq
where $\zeta_k= 1,\ldots,N$, is an index of the mixture in~(\ref{eq:PFSIR}). If we draw from the joint density and then discard the index, then we will have a sample from~(\ref{eq:PFSIR}) as required. This procedure carries out the prediction and update steps of the optimal filter in~(\ref{eq:TMP}). 
We can approximate (\ref{eq:apfd}) by
\beq
\bal
P_\theta(&\Xk,\zeta_k \!\mid\! \Y_{1:k},c)\,\\
&\propto\,p_\theta(\Yk \!\mid\! \mu_{k,\zeta_k},c)\,P_\theta(\Xk\!\mid\!
\x_{k-1,\zeta_k},c)\,w_{k-1,\zeta_k}\,, %\:\: \zeta_k = 1,\ldots,N,
\label{eq:apfd2} 
\eal
\eeq
where $\mu_{k,i}$ is the mode associated with the density of $P_\theta(\X_k \mid \x_{k-1,i},c)$, given by~\cite{ImanBrag:PF}:
\beq
\bal
\mu_{k,i}\,&=\,\text{Mode}[\Xk \mid \x_{k-1,i},\theta,c]\,=\,\f^c_\theta(\x_{k-1,i})\,,
\label{eq-mean}
\eal
\eeq
for $i=1,\ldots,N$, where we have used~(\ref{eq-sgnmodel}) and
the fact that the noise is zero-mode (i.e., the Bernoulli noise intensity $p$ is smaller than $0.5$). 

By simulating the index with the probability $v_{k,i} = p_\theta(\Y_k\mid \mu_{k,i},c) \,w_{k-1,i}$, we can sample from $P_\theta(\Xk,\zeta_k \!\mid\! \Y_{1:k},c)$ and then sample
from the transition density given the mixture, $P_\theta(\X_k \mid \x_{k-1,i},c)$.

Actually, we simulate only from particles associated with large predictive likelihoods. We first sample $N$ times from the joint density of $P_\theta(\Xk,\zeta_k \!\mid\! \Y_{1:k},c)$, and then obtain the new particles $\{\x_{k,i}\}_{i=1}^N$ and their associated
weights $\{w_{k,i}\}_{i=1}^N$
by

\beq
\bal
\x_{k,i}\,&=\,\mu_{k,\zeta_{k,i}}\,\oplus\,\n_{k,i}  \sim P_\theta(\X_k \mid \x_{k-1,\zeta_{k,i}},c), \\
{w}_{k,i}\,&=\,\frac{p_\theta(\Y_k\mid \x_{k,i},c)}{p_\theta(\Y_k\mid\mu_{k,\zeta_{k,i}},c)}\,, \:\: i=1,\dots,N.
\eal
\eeq
% where 
% $\n_{k,i}$ is a realization of process noise at time step $k$.
%\end{enumerate}
The auxiliary variables
 $\{\zeta_{k,i}\}_{i=1}^N$ are obtained by sampling from a discrete distribution:
 \begin{equation}
 \{\zeta_{k,i}\}_{i=1}^N \sim {\rm Cat}(\{v_{k,i}\}_{i=1}^N)\,,
 \end{equation}
where Cat$(a_1, ..., a_N)$ represents a categorical distribution with the probability mass function $f(\zeta = i) = a_i/\sum_{j=1}^N a_j$.

It is shown in~\cite{ImanBrag:PF,pitt1999filtering} that the log-likelihood function in~(\ref{eq:loglik}) can be approximated by:
\beq
\bal
L_c^{\theta}(\Y_{1:T})%&\approx\hat{L}_c^{\theta}(\Y_{1:T})\\
&=p_\theta(\Y_1\mid c)+\sum_{k=2}^T \log p_\theta(\Y_k\mid\Y_{1:k-1},c)\\
&\approx\sum_{k=1}^T\log\left[\left(\frac{1}{N}\,\sum_{i=1}^N v_{k,i}\right)\,\left(\frac{1}{N}\,\sum_{i=1}^N {w}_{k,i}\right)\right]\\
&\triangleq \hat{L}_c^{\theta}(\Y_{1:T}),
\eal
\eeq
where $\hat{L}_c^{\theta}(\Y_{1:T})$ denotes the approximation of $L_c^{\theta}(\Y_{1:T})$. Note that the computational complexity of this algorithm is of order $O(NT)$ which can be much smaller than $O(2^{2n}T)$ that is the complexity of computing the exact log-likelihood function in~(\ref{eq:loglik}).

\begin{figure*}[t]
  \includegraphics[width=\textwidth]{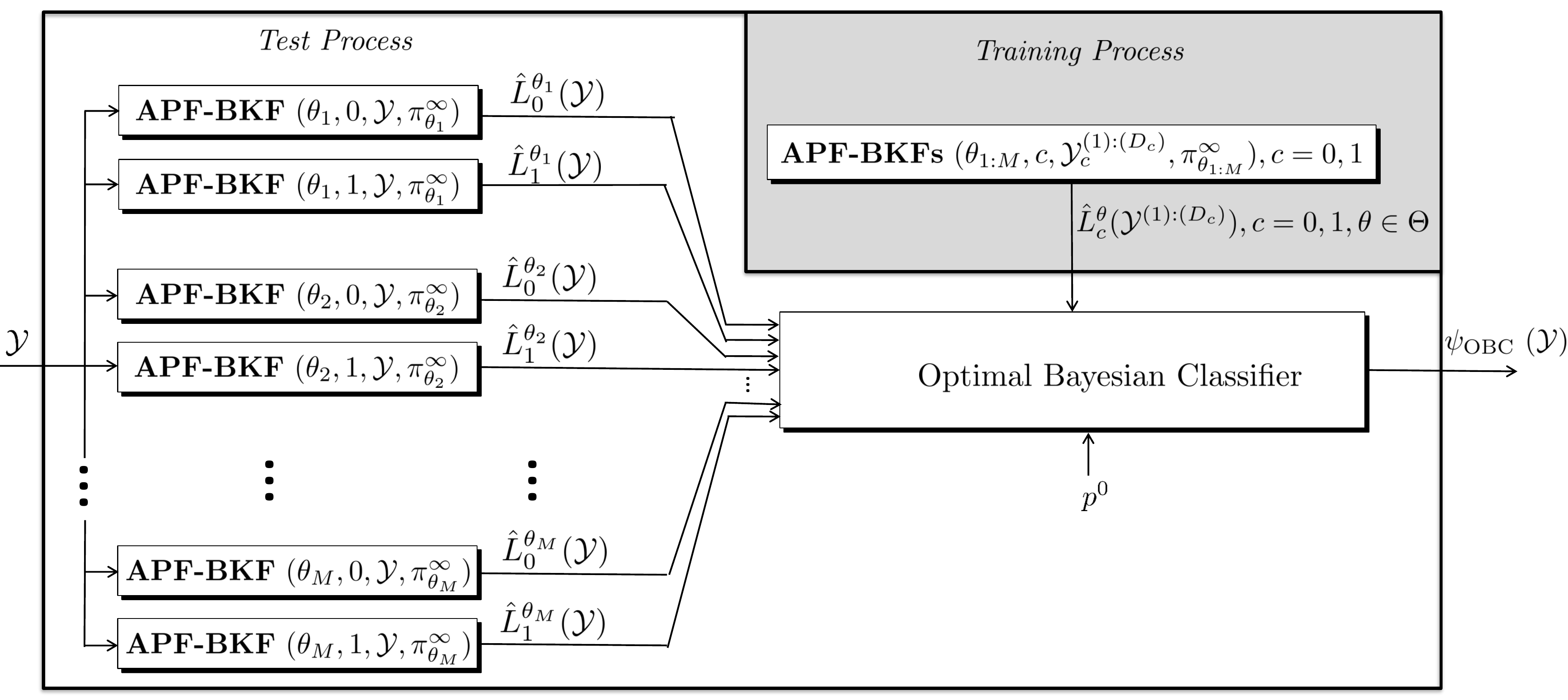}
  \caption{\csentence{The schematic diagram of the proposed method.}}\label{fig:schem}
      \end{figure*}

The whole process and the schematic diagram of the proposed classifier are presented in Algorithm~\ref{alg:PF} and Figure~\ref{fig:schem}, respectively. During the training process, $2M D_0 D_1$ particle filters need to be run for computing the log-likelihood functions of trajectories from all network models in the uncertainty class for two classes. The output values of the particle filters can be used for efficient approximation of the posterior distribution in~(\ref{eq:post}). Then, during the test process, for a given test trajectory, $2M$ particle filters need to be performed for log-likelihood approximation of all network models ($\theta\in\Theta$) and classes ($c=0,1$). These log-likelihood values and posterior probability approximated during the training process can be used to derive the approximate OBC in~(\ref{eq:OBC2}). 

The training process of the proposed method has the computational complexity $O(2 N MT D_0 D_1)$, whereas the exact solution has the complexity $O({2^{2n+1}}MT D_0$ $D_1)$. The exponential growth of the complexity with the size of network (i.e., number of genes) for the exact solution precludes its application to large GRNs. However, the number of particles, $N$, by the proposed method can be chosen relatively small according to the attractor structure of the system (i.e., $N<<2^{2n}$)~\cite{qian2010state}, allowing the classification of large-scale single-cell trajectories (see~Section~\ref{sec:NE}). The complexity of the test process for the proposed method is $O(2NMT)$, as opposed to $O(2^{2n+1}MT)$ for the optimal solution.

\begin{algorithm}[H]
\caption{Scalable Classification of Single-Cell Trajectories}
\begin{algorithmic}[1]
\Statex \underline{\textit{Training Process}}\vspace{1ex}
\State Approximate the steady-state distribution, $\pi_\theta^\infty$, $\theta\in\Theta$ by Monte-Carlo simulations. \vspace{1ex} 
\For {$c\in\{0,1\}$}\vspace{.5ex}
\For {$d\in\{1,...,D_c\}$}\vspace{.5ex}
\For {$\theta\in\Theta$}\vspace{1ex}
\State $\hat{L}_c^\theta(\sY^{(d)}_c)\leftarrow\textbf{APF-BKF}~(\theta,c,\sY^{(d)}_c,\pi_\theta^\infty)$\vspace{1ex}
\EndFor\vspace{.5ex}
\EndFor\vspace{.5ex}
\EndFor\vspace{1ex}
\Statex \underline{\textit{Test Process}}\vspace{1ex}
\For {$c\in\{0,1\}$}\vspace{.5ex}
\For {$\theta\in\Theta$}\vspace{1ex}
\State $\hat{L}_c^\theta(\sY)\leftarrow\textbf{APF-BKF}~(\theta,c,\sY,\pi_\theta^\infty)$\vspace{1ex}
\EndFor\vspace{1ex}
\State $\tau_\Theta^c(\sY)=\sum_{\theta\in\Theta}\frac{\exp(\sum_{d=1}^{D_c}\hat{L}_c^\theta(\sY^{(d)}_c))\,\pi(\theta\mid c)\exp(\hat{L}_c^\theta(\sY))}{\sum_{\theta'\in\Theta}\exp(\sum_{d=1}^{D_c}\hat{L}_c^{\theta'}(\sY^{(d)}_c))\,\pi(\theta'\mid c)}$\vspace{1ex}
\EndFor\vspace{.5ex}
\State $\psi_{\rm OBC}(\sY)\,=\,\begin{cases}
0 & \text{if } p^0 \,\tau^0_\Theta(\sY) \geq (1-p^0) \,\tau^1_\Theta(\sY) \\
1 & \text{otherwise}
\end{cases}$. \vspace{1ex}
\end{algorithmic}
\begin{algorithmic}[1]
\hrule\vspace{1ex}
\Statex \hspace{-3ex}{\bf APF-BKF}~($\theta,c,\Y_{1:T},\pi_{0|0}$)~\cite{ImanBrag:PF}\vspace{1ex}
%\hrule\vspace{1ex}
\State $\hat{L}_{c}^\theta=0, \x_{0,i} \sim {\pi}_{0|0}, w_{0,i}=1/N$, for $i=1,\ldots,N$.\vspace{1ex}
\For{$k =1,2,\ldots,$} \vspace{1ex}
% \For{$i =1$ to $N$}\vspace{0.8ex}
\State $\mu_{k,i}\,=\,\f_\theta^c(\x_{k-1,i})$, $i=1,...,N$.\vspace{1ex}
\State ${v}_{k,i}\,=\,p_\theta(\Y_k\mid\mu_{k,i},c)\,w_{k-1,i}$, $i=1,...,N$.\vspace{1ex}
\State $\{\zeta_{k,i}\}_{i=1}^N \sim {\rm Cat}(\{v_{k,i}\}_{i=1}^N)$.\vspace{1ex}
%\For{$i =1$ to $N$}\vspace{0.8ex}
\State $\x_{k,i}\,=\,\mu_{k,\zeta_{k,i}}\oplus \n_{k,i}$, $i=1,...,N$.\vspace{1ex}
\State ${w}_{k,i}\,=\,\frac{p_\theta(\Yk\mid\x_{k,i},c)}{p_\theta(\Yk\mid\mu_{k,\zeta_{k,i}},c)}$, $i=1,...,N$.\vspace{1ex}
 \State $\hat{L}_{c}^\theta=\hat{L}_{c}^\theta+\log\left[\left(\frac{1}{N} \sum_{i=1}^N v_{k,i}\right)\left(\frac{1}{N} \sum_{i=1}^N {w}_{k,i}\right)\right]$.\vspace{1ex}
\State $w_{k,i}={w}_{k,i}/\sum_{j=1}^N {w}_{k,j},\,i=1,\ldots,N$.\vspace{1ex}
\EndFor\vspace{.8ex}
\Statex \hspace{-3ex} {\bf Return} ($\hat{L}_c^\theta$)
\end{algorithmic}\label{alg:PF}
\end{algorithm}

\subsection*{Optimal Bayesian Classifier for Multiple-Cell Scenarios}
In the previous sections, the classification of single-cell trajectories is discussed. Here, we consider common scenarios in molecular biology research where gene-expression data are often based on the average expression from multiple-cells at different time with different states. Since the trajectories are assumed to be independent and drawn based on the dynamics of the true network, its steady-state distribution $\pi^\infty_{\theta^*}$ characterizes the probability of the system being at different states. It can be shown that the  multiple-cell data are independent samples from the following measurement model \cite{Alireza:TCBB2}:
\beq\label{eq:mul}
\Y \sim \N(\boldsymbol{\lambda
}+\sum_{i=1}^{2^n} D\x^i\pi_{\theta^*,c}^\infty(i), \sigma^2)\,,
\eeq
where $\{\x^1,...,\x^{2^n}\}$ denotes all the network states in the Boolean vector representation. However, we assume that the true network model $\theta^*$, is unknown, and is represented by a finite set of $M$ possible network models $\{\theta_1,...,\theta_M\}$ with prior probability $\pi(\theta\mid c)$. Let $\Y_c^{(1)},...,\Y_c^{(N_c)}$ be the multiple-cell training measurements available for class $c$. The optimal Bayesian classifier for a given test sample $\Y$ can be represented by
\beq\label{eq:OBC3}
\psi_{\rm OBC}(\Y)=\begin{cases}
0 & \text{if } p^0 E_{\theta\mid\Y_0^{(1)},...,\Y_0^{(N_0)},c=0}[p_\theta(\Y\mid c=0)] \\
&\geq(1-p^0) \\
&\quad\times E_{\theta\mid\Y_1^{(1)},...,\Y_1^{(N_1)},c=1}[p_\theta(\Y\mid c=1)].\\
1 & \text{otherwise}
\end{cases}
\eeq
The posterior probability of the parameter $\theta$ can be computed as
\beq\label{eq:likmu}
\bal
&P\left(\theta\mid\Y_c^{(1)},...,\Y_c^{(N_c)}\right)\\
&\quad\qquad=\,\frac{p_\theta(\Y_c^{(1)},...,\Y_c^{(N_c)}\mid c)\,\pi(\theta\mid c)}{\sum_{\theta'\in\Theta}p_{\theta'}(\Y_c^{(1)},...,\Y_c^{(N_c)}\mid c)\,\pi(\theta'\mid c)}\,\\
&\quad\qquad=\,\frac{\left(\prod_{d=1}^{N_c} p_\theta(\Y_c^{(d)}\mid c)\right)\,\pi(\theta\mid c)}{\sum_{\theta'\in\Theta} \left(\prod_{d=1}^{N_c} p_{\theta'}(\Y_c^{(d)}\mid c)\right)\,\pi(\theta'\mid c)}.
\eal
\eeq
Computation of (\ref{eq:OBC3}) requires computing the conditional probability of training and test samples given all $\theta\in\Theta$ and $c\in\{0,1\}$. Let ${\bf A}$ be an $n\times 2^n$ matrix containing all Boolean states of the system (i.e., ${\bf A}=[\x^1,...,\x^{2^n}]$). According to~(\ref{eq:mul}), %\hl{again, this is not right, you are only looking gene j?}
the conditional distribution of an arbitrary sample $\tilde{\Y}$ given class $c$ and network model $\theta$ can be written as
\beq\label{eq:likmul}
\bal
\tilde{\Y}\mid c, \theta\,\sim&\,\\
 &(1- %\sum_{i=1}^{2^n} \x^i
{\bf A}\,\pi^\infty_{\theta,c})\,\circ \N(\boldsymbol{\lambda}, \sigma^2 {\bf I}_n) \\
&+ 
%\sum_{i=1}^{2^n} \x^i
({\bf A}\,\pi^\infty_{\theta,c})\,\circ\,\N(\boldsymbol{\lambda}+D{\bf 1}_n, \sigma^2 {\bf I}_n). 
\eal
\eeq
where ``$\circ$" is the Hadamard product. 
Thus, the Boolean nature of the state vector suggests that each element of the multiple-cell measurement is distributed as a mixture of two Gaussian distributions. Replacing (\ref{eq:likmu}) and (\ref{eq:likmul}) into (\ref{eq:OBC3}) leads to the OBC for multiple-cell scenarios. Comprehensive comparison results between the OBC in single-cell trajectories and multiple-cells are provided in the next section.

\section*{Results and Discussion}
\label{sec:NE}

\begin{figure}[t]
\includegraphics[width=.4\textwidth]{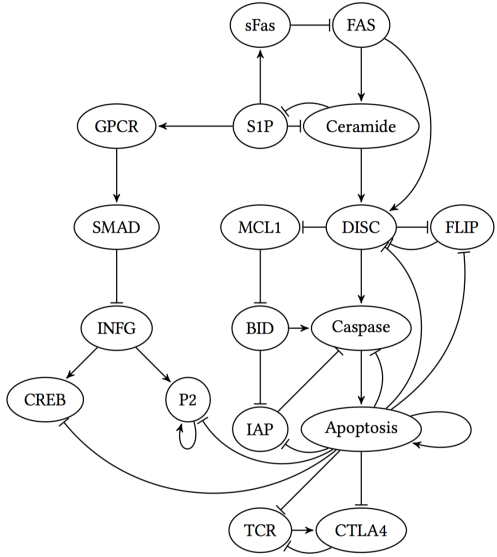}
  \caption{\csentence{T-LGL leukemia gene regulatory network.}
      }\label{GRN}
\end{figure}

We evaluate the proposed single-cell trajectory classifier and compare its performance with the OBC based on multiple-cell average expression on the T-LGL leukemia Boolean network, whose GRN \cite{saadatpour2011dynamical} is shown
in Figure~\ref{GRN}. This GRN has $18$ genes. The regulating functions
are defined in Table~\ref{table:booleanRules} \cite{saadatpour2011dynamical, zhang2008network}. The main node is ``Apoptosis'', which denotes programmed cell death. According to \cite{saadatpour2011dynamical}, in the mutated
case, the node Apoptosis is stuck at OFF state and cannot be activated. As a result, we derive the healthy Boolean network from Table~\ref{table:booleanRules} and for the mutated Boolean network we put the value of Apoptosis in Table~\ref{table:booleanRules} to zero. This means that in the cancerous scenario, the value of Apoptosis does not obey the regulating functions and is always zero.

\begin{table}[t]
  \caption{Definitions of Boolean functions for the T-LGL leukemia Boolean network with 18 nodes \cite{saadatpour2011dynamical, zhang2008network}.}
  \label{table:booleanRules}\vspace{1mm}
  \centering
  \begin{tabular}{ll}
    \toprule
    Node     & Regulating function\\
    \midrule
    CTLA4 &  TCR $\wedge$ $\neg$ Apoptosis  \\
    TCR &  $\neg$ (CTLA4 $\vee $ Apoptosis) \\
    CREB &  IFNG $\wedge$ $\neg$ Apoptosis \\
    IFNG & $\neg$ (SMAD $\vee $ P2 $\vee $ Apoptosis) \\
    P2 & (IFNG $\vee $ P2) $\wedge$ $\neg$ Apoptosis \\
    GPCR & S1P $\wedge$ $\neg$ Apoptosis \\
    SMAD & GPCR $\wedge$ $\neg$ Apoptosis\\
    Fas & $\neg$ (sFas $\vee $ Apoptosis)\\
    sFas & S1P $\wedge$ $\neg$ Apoptosis\\
    Ceramide & Fas $\wedge$ $\neg$ (S1P or Apoptosis)\\
    DISC & (Ceramide $\vee $ (Fas $\wedge$ $\neg$ FLIP)) $\wedge$ $\neg$ Apoptosis\\
    Caspase & ((BID $\wedge$ $\neg$ IAP) $\vee $ DISC) $\wedge$ $\neg$ Apoptosis\\
    FLIP & $\neg$ (DISC $\vee $ Apoptosis)\\
    BID & $\neg$ (MCL1 $\vee $ Apoptosis)\\
    IAP & $\neg$ (BID $\vee $ Apoptosis)\\
    MCL1 & $\neg$ (DISC $\vee $ Apoptosis)\\
    S1P & $\neg$ (Ceramide $\vee $ Apoptosis)\\
    Apoptosis & Caspase $\vee $ Apoptosis \\
    \bottomrule
  \end{tabular}
\end{table}

As we may not know the true network function, we consider four candidate network functions for each of the healthy and mutated networks as the uncertainty class of possible GRN models. In addition to the true network, we remove the operation $\neg$ of Apoptosis for the genes sFas and GPCR, which are intermediate nodes. Therefore, this network is very close to the true network. For the third network, we remove $\neg$ of Apoptosis from two other nodes IAP and P2. In the fourth network of this uncertainty class, we change the operation AND to OR for the gene BID. In this study, we use the observation models described in equations~(\ref{eq:obs}) and~(\ref{eq:mul}) for the single-cell trajectory and multiple-cell averaging, respectively.

\begin{figure*}[t]
\includegraphics[width=\textwidth]{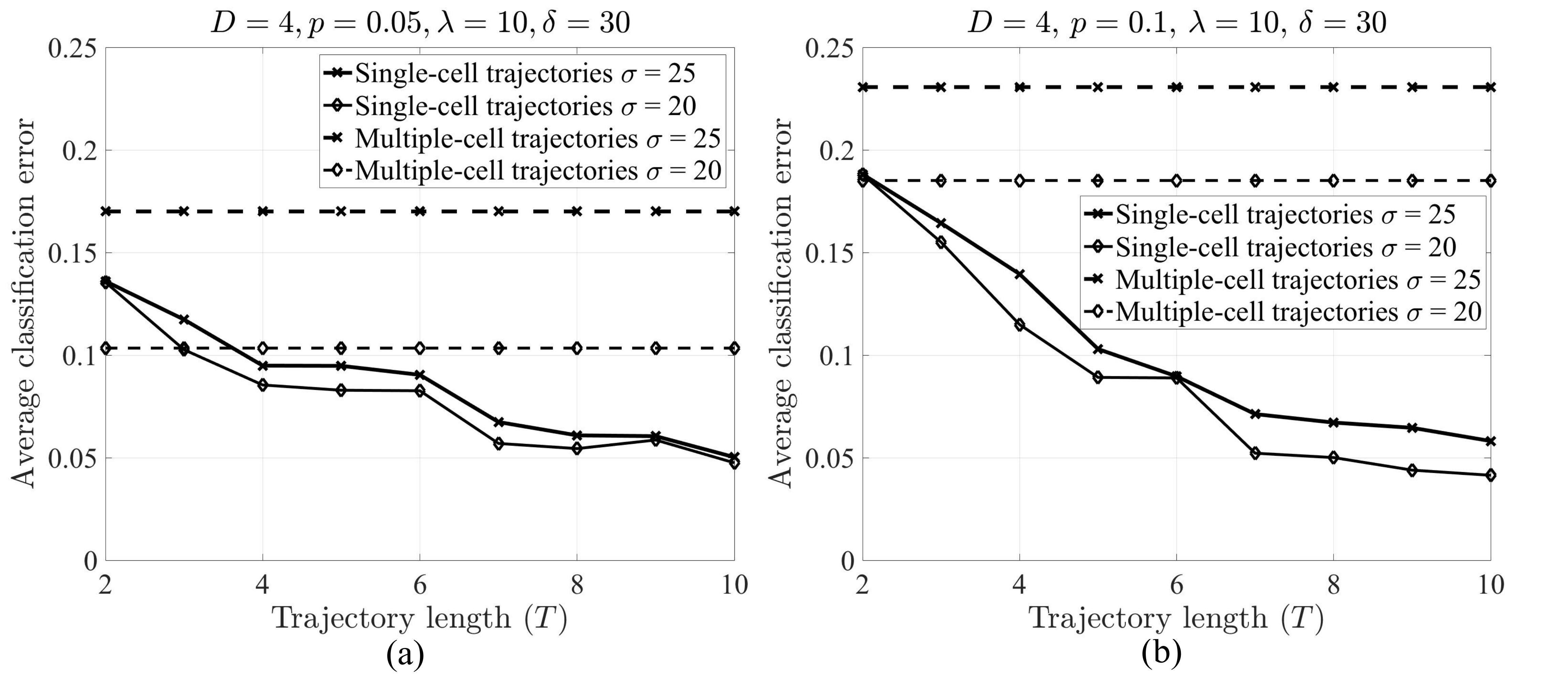}
  \caption{\csentence{Classification errors using the trajectory and multiple-cell classifiers in the T-LGL leukemia Boolean network.} (a) The parameter is $p = 0.05$, (b) The parameter is $p = 0.1$.}\label{fig:vsT}
      \end{figure*}

Figures~\ref{fig:vsT}(a) and~\ref{fig:vsT}(b) show the classification errors for the simulated data based on the T-LGL leukemia Boolean network versus the number of time points $T$ for two values of state perturbation probability $p=0.05$ and $0.1$, respectively. For the sake of simplicity, we assume the  gene-expression parameters in equations~(\ref{eq:obs}) and~(\ref{eq:mul}) to be the same for all genes, that are $\boldsymbol{\lambda}=[\lambda,...,\lambda]^T$ and $D=\text{Diag}(\delta,...,\delta)$, and set $\lambda = 10$, and $\delta = 30$. Two different values are considered for the observation noise level: $\sigma = 20$ (low
noise), and $\sigma = 25$ (high noise). While $\sigma$ corresponds to both within subject and between subject variations in the single-cell, it shows between subject variation in the multiple-cell because multiple-cells would allow to average out the within subject variance.
We set the number of training trajectories $D=D^{c=0} + D^{c=1}=4$. In both figures, the error curves are monotonically decreasing in terms of the trajectory length $T$ for the single-cell classifier. There is a special case in which the error gets fixed after some $T$. This may be explained by the effect from the steady-state distributions depending on the lengths of attractor cycles of the networks under study. 
When the perturbation noise $p$ and the observation noise $\sigma$ are small, the sufficient $T$ to achieve the least possible error is $L + 1$, where $L$ is the minimum attractor length in the two networks. More precisely, the BNps tend to the deterministic BNs when the perturbation noise is small, meaning that the observations occur only in
the attractor states and circulate inside the attractor cycles. In such a case, $L + 1$ is the maximum length of a trajectory that can help distinguish the two networks.
But there is a nonnegligible probability of jumping states in the considerable perturbation scenarios, so that longer trajectories can be helpful. In all figures for every value of $T$ and $p$, the error increases with increasing observation noise. While the proposed classifier works in both low- and high-noise scenarios, the classifier based on the multiple-cell expression data only works well in the low-noise scenarios and is very sensitive to $\sigma$. In low observation noise scenarios and when $p = 0.05$, multiple-cell classifier can easily classify, while the trajectory-based with two time-points will have a little bit higher error due to the common short segments of the trajectories between two classes. When there are at least four time-points, which is the attractor cycle size in the uncertainty class of the networks, the performance of the classifiers based on single-cell trajectories is better compared to the ones using multiple-cell even in the low-noise scenarios. Increasing the number of time-points help better decipher the difference in single-cell trajectories between two classes (healthy vs. cancerous) with improved classification accuracy. Compared to the multiple-cell classifier based on averaged gene expression over cells at different states, the trajectory based classifier can clearly improve the classification performance. Even using four time points, the classification accuracy can be improved up to 8\%. Using longer trajectories, the improvement can be up to 18\% as indicated in Figure~\ref{fig:vsT}.

\begin{figure}[t]
\includegraphics[width=.4 \textwidth]{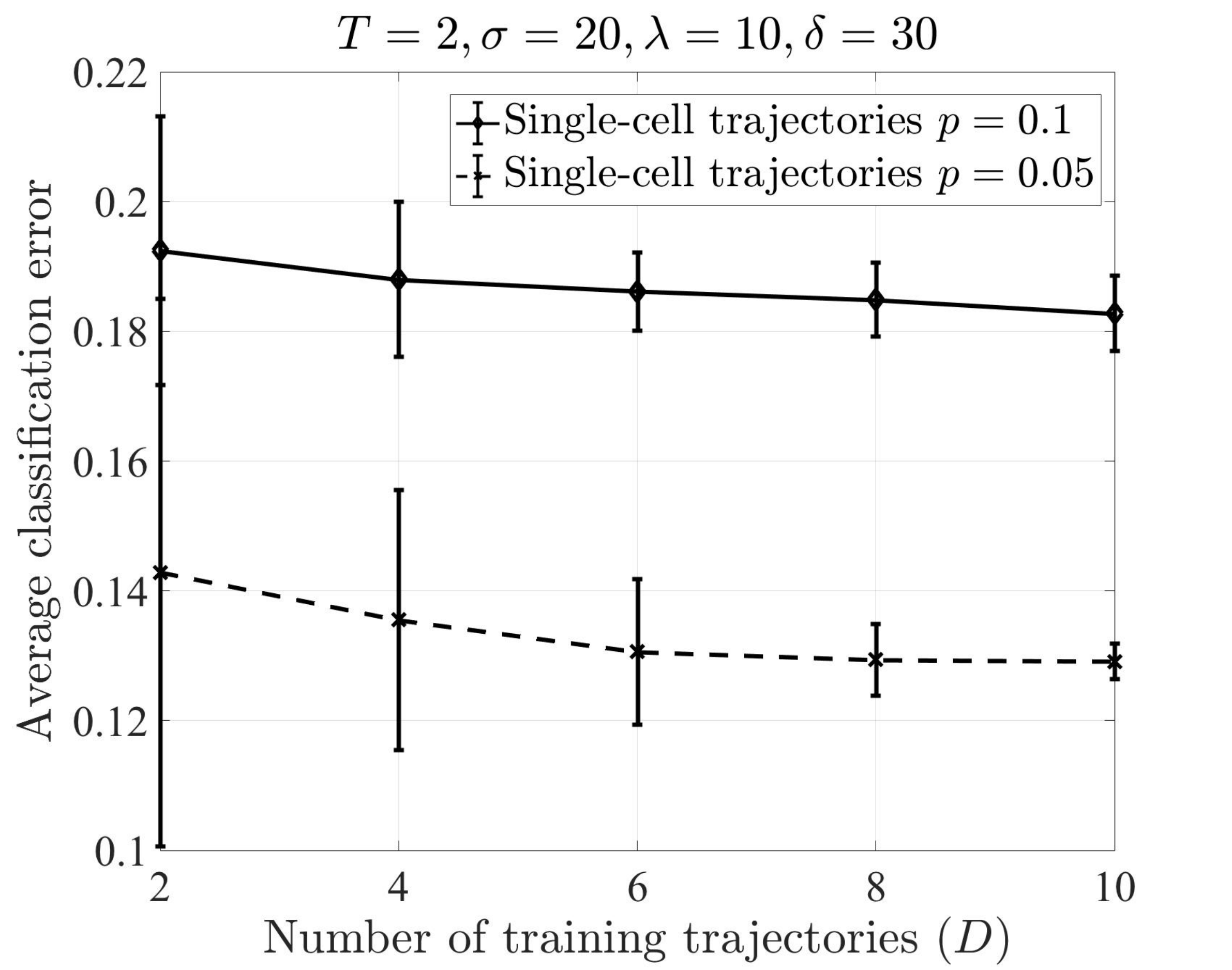}
  \caption{\csentence{Classification error of the single-cell classifier versus training sample size for the T-LGL leukemia Boolean network.}
      } \label{fig:Ntrain}
      \end{figure}

In addition to the effect of trajectory length, we would like to investigate how the number of training trajectories affect the performance of the proposed method, especially with a low number of training samples.
Figure~\ref{fig:Ntrain} shows the effect of the number of training trajectories $D$ on the classification performance.
In the particle filter point of view, increasing $D$ by 1 means increasing the available data as $T$. Therefore, we set $T=2$, that is smallest $T$, to better see the trend of classification error. Both the average error and its standard deviation decrease with more training trajectories and the classification error converges to a fixed value when $D$ becomes large enough. The value of $D$ required for a converged error rate depends on the parameters $T$, $p$, and $M$. In real-world scenarios that may have significant uncertainty and the perturbation probability is high, we need more training data to improve the performance. 

\begin{figure}[t]
\includegraphics[width=.4\textwidth]{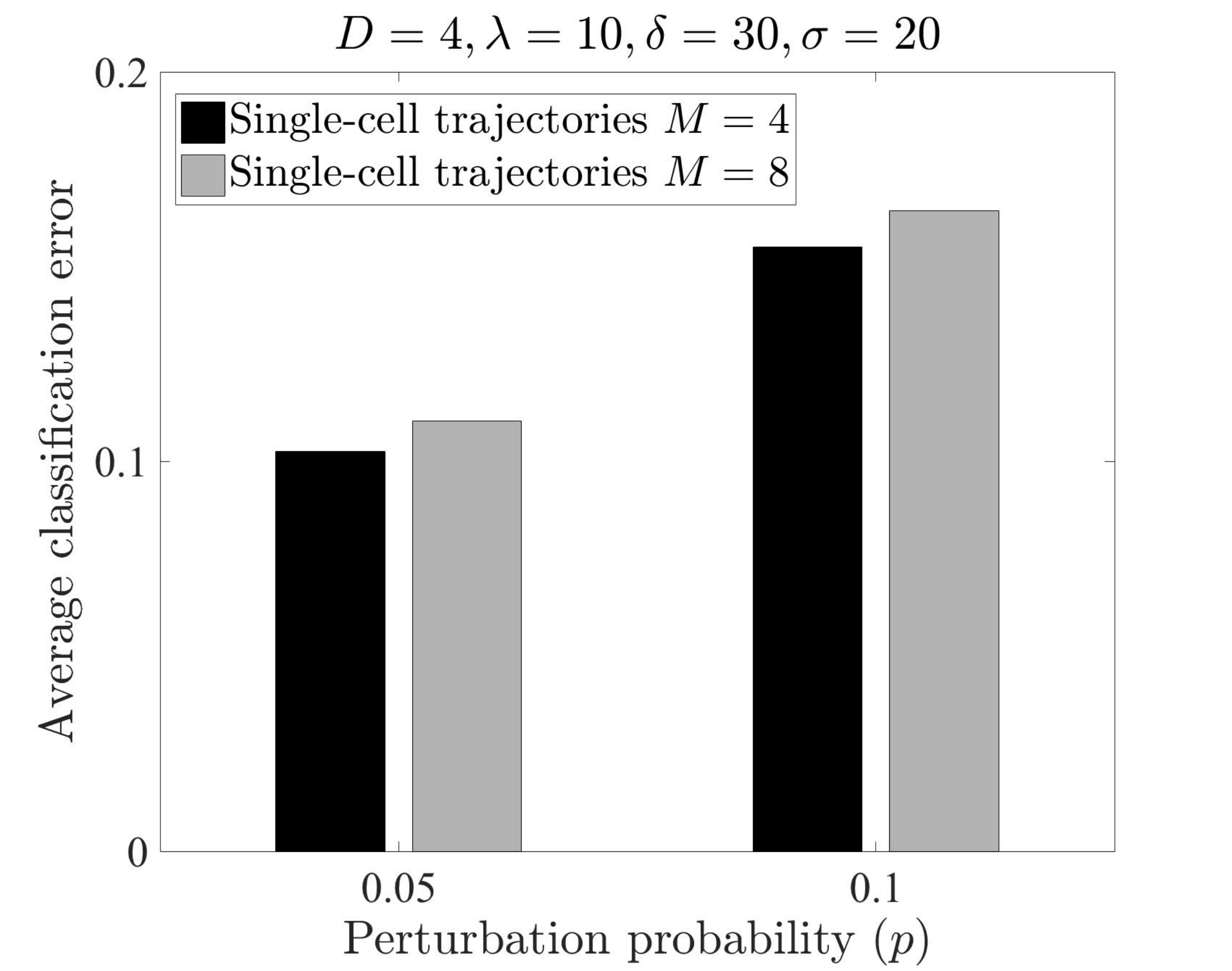}
  \caption{\csentence{Classification errors using the single trajectory classifier in the T-LGL leukemia Boolean network versus the number of uncertain networks.}
      }\label{fig:unc}
      \end{figure}
\begin{figure}[t]
\includegraphics[width=.4\textwidth]{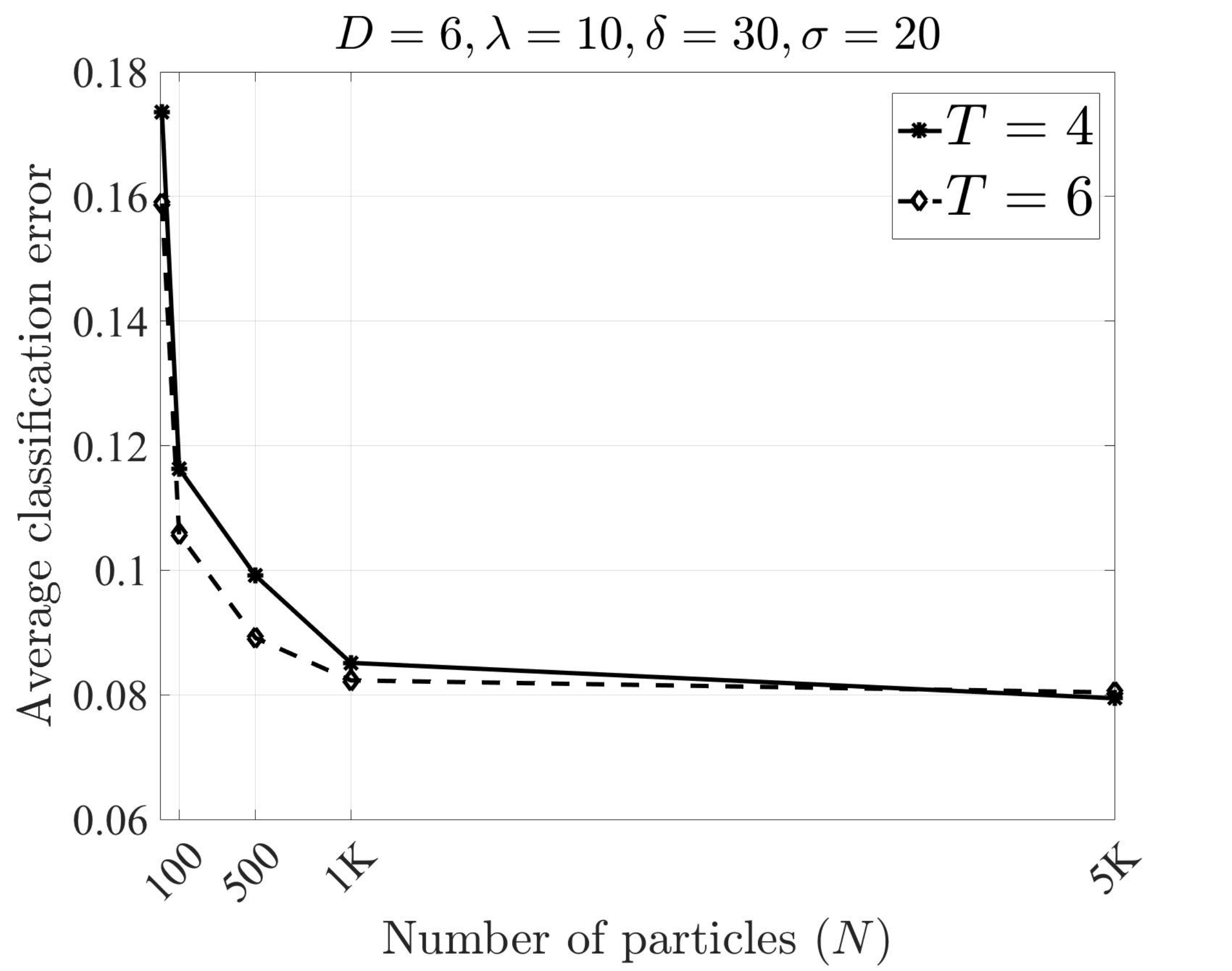}
  \caption{\csentence{Classification errors using the trajectory based classifier in the T-LGL leukemia Boolean network versus the number of particles.}
      }\label{fig:numberofpf}
      \end{figure}

Figure~\ref{fig:unc} illustrates how the model uncertainty may affect the classification performance. In our setup, the uncertainty is manifested as the size of the network uncertainty class -- the number of uncertain networks for each class. In both perturbation probabilities, the error increases with increasing $M$. 
To demonstrate how the proposed method can reduce the computational cost of the Boolean network classification, we check the change of the average classification error with respect to the number of particles. Figure~\ref{fig:numberofpf} shows that increasing the number of particles monotonically decreases the classification error and the average error converges with only 1,000 particles. This can be compared with the  traditional method \cite{Alireza:TCBB1}, which needs computation based on the $2^{18} \times 2^{18}$ transition probability matrix. Such a dimension is too large for the direct application of the OBC. 

To more comprehensively evaluate the proposed method, we have compared it with two other classification methods, i.e. IRB \cite{Alireza:BMC} and Plug-In \cite{Alireza:TCBB2}. The performance comparisons under four different scenarios are provided in Table~\ref{table:comsigma25}. The OBC stands out as the best performing method in different scenarios. Using OBC improves the accuracy by 8.5\% and 1.5\% with $T=3$ for $p=0.1$ and $\sigma =25$ compared to Plug-In and IRB, respectively.

\begin{table}[h]
  \caption{Trajectory based classification results for high-noise scenario ($\sigma = 25$).}
  \label{table:comsigma25}\vspace{1mm}
  \centering
  \begin{tabular}{lllll}
    \toprule
    \multicolumn{4}{c}{$\qquad\qquad\quad \mbox{(}p =0.05 \mbox{)}$}{$\qquad$ ($p = 0.1$)}                   \\
    \cmidrule(r){2-5}
    Method     & $T = 3$     & $T = 7$ & $T = 3$ & $T = 7$\\
    \midrule
    Plug-In & $0.1777$  & $0.0922$    &  $0.2524$ & $0.1774$ \\
    IBR     &  $0.1384$ &  $0.0723$  & $0.1800$  & $0.0750$  \\
    OBC     & $\mathbf{0.1173}$  & $\mathbf{0.0674}$    &  $\mathbf{0.1643}$ & $\mathbf{0.0646}$ \\
    \bottomrule
  \end{tabular}
\end{table}

\tce{The proposed method does not have any restriction on the noise distribution assumptions due to the generalizability of particle filters. To show this, we also test our method with different noise distributions. While the noise of GRNs is usually Gaussian, sometimes due to up regulation, noise can be Poisson or Negative Binomial (NB). We have simulated Gaussian and NB noise distributions with the same mean and variance while for Poisson noise, the variance is equal to its mean (Details can be found in Additional file 1, the additional experimental results). Additional File 1, Figure~S1 shows that our method performs consistently well for all observation noises. The Poisson results are superior to the other models as its variance is smaller.}

We also compare the performance of APF with the plain Sequential Importance Resampling (SIR) in high process noise. As Additional File 1, Figure~S2 shows, the performance is similar especially when there is enough time points. When the number of time points is low, the SIR-based particle filter performs worse. The superior performance of APF in real-world GRNs is due to the fact that the size of attractors is usually small and the predicted mode values by APF can be a good approximation for the next prediction. Moreover, the initial distribution is assumed to be from the stationary distribution. This makes APF a more desirable approximation solutions due to the lower diversity in the particles.

\section*{Conclusions}
In this paper, we have developed the optimal Bayesian classifier for binary classification of single-cell trajectories under regulatory model uncertainty. The partially-observed Boolean dynamical system is used for modeling the dynamical behavior of gene regulatory networks. Due to the intractability of the OBC for large GRNs, we have proposed a particle filtering technique for approximating the OBC. This particle-based solution reduces the computational and memory complexity of the optimal solution significantly. The performance of the proposed particle-based method is demonstrated through numerical experiments using a POBDS model of the well-known T-cell large granular lymphocyte (T-LGL) leukemia network based on noisy time-series gene-expression data.

\section*{Abbreviations}
GRN: Gene regulatory network; OBC: Optimal Bayesian classifier; IRB: Intrinsically Bayesian robust; POBDS: Partially observed Boolean dynamical system; BKF: Boolean Kalman filter; BKS: Boolean Kalman smoother;  APF-BKF:  Auxiliary  particle-filter  implementations of  the Boolean  Kalman  filter; SMC: Sequential Monte-Carlo; BNp: Boolean network with perturbation; MMSE:  Minimum mean square error; T-LGL: T-cell large granular lymphocyte.

%%%%%%%%%%%%%%%%%%%%%%%%%%%%%%%%%%%%%%%%%%%%%%
%%                                          %%
%% Backmatter begins here                   %%
%%                                          %%
%%%%%%%%%%%%%%%%%%%%%%%%%%%%%%%%%%%%%%%%%%%%%%

\begin{backmatter}

\section*{Declarations}

\section*{Competing interests}
  The authors declare that they have no competing interests.

\section*{Author's contributions}
E. H. developed OBC for the single-cell trajectory classification, developed the scalable OBC, performed the experiments, and wrote the manuscript.
M. I. wrote part of the code and the first draft, and in conjunction with U. B. N. structured the APF-BKF by integrating their previous partially-observed Boolean dynamical system into this new framework.
E. R. D. and X.Q. oversaw the project, proposed the new scalable OBC, and wrote the manuscript. All authors have read and approved the final manuscript.

\section*{Availability of data and materials}
Not applicable.

\section*{Funding}
This research was funded in part by a grant from the Brookhaven National Laboratory (BNL) and NSF Awards CCF-1553281 and CCF-1718513. Publication costs are funded in part by a grant from the Brookhaven National Laboratory (BNL) and NSF Award CCF-1718513. 

\section*{Ethics approval and consent to participate}
Not applicable.

\section*{Consent for publication}
Not applicable.

\section*{Acknowledgment}
We thank Texas A\&M High Performance Research
Computing and Texas Advanced Computing Center for
providing computational resources to perform experiments in
this work.

\bibliographystyle{bmc-mathphys} % Style BST file (bmc-mathphys, vancouver, spbasic).
\bibliography{ref.bbl}      % Bibliography file (usually '*.bib' )

\section*{Figures}

%%%%%%%%%%%%%%%%%%%%%%%%%%%%%%%%%%%
%%                               %%
%% Tables                        %%
%%                               %%
%%%%%%%%%%%%%%%%%%%%%%%%%%%%%%%%%%%

%% Use of \listoftables is discouraged.
%%
% \section*{Tables}

% \setcounter{table}{0}

%%%%%%%%%%%%%%%%%%%%%%%%%%%%%%%%%%%
%%                               %%
%% Additional Files              %%
%%                               %%
%%%%%%%%%%%%%%%%%%%%%%%%%%%%%%%%%%%

\section*{Additional Files}
\subsection*{Additional file 1 --- Supplemental Materials.}
This additional file contains the additional experiment results.

\end{backmatter}

\end{document}

\maketitle

This document contains the supplementary materials for the paper ``Scalable Optimal Bayesian Classification of Single-Cell Trajectories under Regulatory Model Uncertainty''.

\section*{Additional Experimental Results}

\subsection*{}

\begin{figure}[h]
    \centering
    \includegraphics[width=100mm]{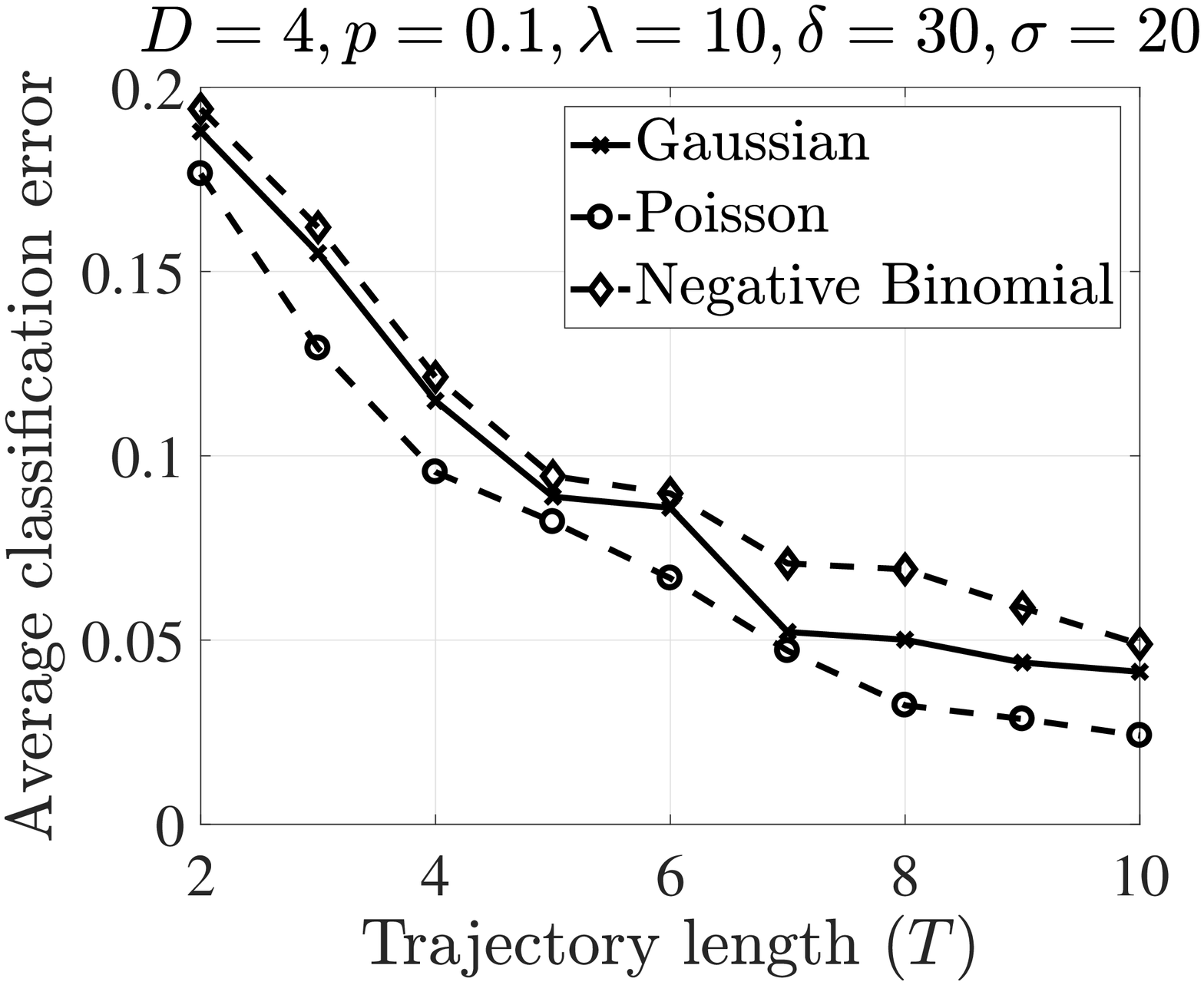}
    \caption{Trajectory-based classification results for different noise distributions.}
    \label{fig:noise}
\end{figure}

In addition to the results for testing different Gaussian noise levels in the main text, we provide more detailed experimental results for different noise distributions as gene expression noise based on the gene regulatory network (GRN) model can also be Poisson or Negative Binomial (NB).
Specifically, Let $\Yk$ denote the expression value at time $k$. To compare different noise models, the observation models as discussed in the main text are set to $(\Yk | \X_k, \boldsymbol{\lambda}, D) \sim \N(\boldsymbol{\lambda}\,+\,D\X_k, \sigma^2I_n)$, $(\Yk | \X_k, \boldsymbol{\lambda}, D) \sim \mathrm{NB}(\boldsymbol{r}_{\mathrm{NB}}, \boldsymbol{p}_{\mathrm{NB}})$, and $(\Yk | \X_k, \boldsymbol{\lambda}, D) \sim \mathrm{Poisson}(\boldsymbol{\lambda}\,+\,D\X_k)$ for Gaussian, Negative Binomial, and Poisson noise distributions, respectively. For NB parameters,  $\boldsymbol{r}_{\mathrm{NB}} = \frac{(\boldsymbol{\lambda}\,+\,D\X_k)^2}{\sigma^2I_n - (\boldsymbol{\lambda}\,+\,D\X_k)}$ and $\boldsymbol{p}_{\mathrm{NB}} = \frac{\sigma^2I_n - (\boldsymbol{\lambda}\,+\,D\X_k)}{\sigma^2I_n}$.
For the sake of simplicity, we assume the gene-expression parameters to be the same for all genes and $\boldsymbol{\lambda}=[\lambda,...,\lambda]^T$ and $D=\text{Diag}(\delta,...,\delta)$, with $\lambda = 10$, $\delta = 30$, and $\sigma = 20$.
Figure~\ref{fig:noise} shows that the proposed particle filter based method can work consistently well in different noise distributions.
This is mostly due to the generalizability of  particle filter. 

Regarding the performance of auxiliary particle filter (APF) in high process noise, we compared it with the plain Sequential Important Resampling (SIR) as well. As Figure~\ref{fig:sir} shows, the performance is similar especially when there is enough time points. When the number of time points is low, the SIR based particle filter has the worse performance. Based on this comparison, we decided to focus on the APF method. The superior performance is mostly due to the more efficient resampling step of APF which is based on the predicted modes of particles in the next time step, while the random resampling process of SIR algorithm poses the significant errors during the estimation process.

\begin{figure}[h]
    \centering
    \includegraphics[width=100mm]{additionalFile1.eps}
    \caption{Comparison between SIR-based and auxiliary particle filters for different $p$ in high noise scenario.}
    \label{fig:sir}
\end{figure}